\documentclass[prb,twocolumn,showpacs]{revtex4-1}
\usepackage[pdftex]{graphicx}
\usepackage{mathrsfs}
\usepackage[sort&compress]{natbib}

\begin{document}
\title{Evidence for a Spin Phase Transition at $\nu=0$ in Bilayer Graphene}
\author{P. Maher$^{1}$}
\author{C. R. Dean$^{2,3}$}
\author{A. F. Young$^{1}$}
\author{T. Taniguchi$^{4}$}
\author{K. Watanabe$^{4}$}
\author{K. L. Shepard$^{2}$}
\author{J. Hone$^{3}$}
\author{P. Kim$^{1}$}

\affiliation{$^{1}$Department of Physics, Columbia University, New York, NY 10027, USA}
\affiliation{$^{2}$Department of Electrical Engineering, Columbia University, New York, NY, 10027, USA}
\affiliation{$^{3}$Department of Mechanical Engineering, Columbia University, New York, NY, 10027, USA}
\affiliation{$^{4}$National Institute for Materials Science, 1-1 Namiki, Tsukuba, 305-0044, Japan}

\begin{abstract}
The most celebrated property of the quantum spin Hall effect is the presence of spin-polarized counter-propagating edge states\cite{kane_z_2_2005, bernevig_quantum_2006, konig_quantum_2007}. This novel edge state configuration has also been predicted to occur in graphene when spin-split electron- and hole-like Landau levels are forced to cross at the edge of the sample\cite{abanin_spin-filtered_2006, fertig_luttinger_2006, jung_theory_2009, kharitonov_edge_2012}. In particular, a quantum spin Hall analogue has been predicted at $\nu=0$ in bilayer graphene if the ground state is a spin ferromagnet\cite{kharitonov_canted_2012, zhang_distinguishing_2012}. Previous studies have demonstrated that the bilayer $\nu=0$ state is an insulator in a perpendicular magnetic field\cite{feldman_broken-symmetry_2009, zhao_symmetry_2010, weitz_broken-symmetry_2010, kim_spin-polarized_2011, velasco_jr_transport_2012, veligura_transport_2012, freitag_spontaneously_2012}, though the exact nature of this state has not been identified. Here we present measurements of the $\nu=0$ state in a dual-gated bilayer graphene device in tilted magnetic field. The application of an in-plane magnetic field and perpendicular electric field allows us to map out a full phase diagram of the $\nu=0$ state as a function of experimentally tunable parameters. At large in-plane magnetic field we observe a quantum phase transition to a metallic state with conductance of order $4e^2/h$, consistent with predictions for the ferromagnet.
\end{abstract}


\maketitle

Under a strong perpendicular magnetic field, bilayer graphene (BLG) develops a $\nu=0$ quantum Hall (QH) state at the charge neutrality point (CNP) which displays anomalous insulating behavior\cite{feldman_broken-symmetry_2009, zhao_symmetry_2010, weitz_broken-symmetry_2010, kim_spin-polarized_2011, velasco_jr_transport_2012, veligura_transport_2012, freitag_spontaneously_2012}. Transport studies in a dual-gated geometry\cite{weitz_broken-symmetry_2010, kim_spin-polarized_2011, velasco_jr_transport_2012} indicate that this gapped state results from an interaction-driven spontaneous symmetry breaking in the valley-spin space\cite{gorbar_dynamics_2010}. However, the exact order of the resulting ground state remains controversial. In high-mobility suspended BLG devices, a broken symmetry state at charge neutrality has also been observed at zero magnetic field whose nature has been under intense theoretical \cite{castro_low-density_2008, lemonik_spontaneous_2010, vafek_many-body_2010, nandkishore_quantum_2010, kharitonov_correlated_2011, zhang_distinguishing_2012} and experimental \cite{mayorov_interaction-driven_2011, veligura_transport_2012, weitz_broken-symmetry_2010, velasco_jr_transport_2012, freitag_spontaneously_2012} investigation. Some experiments indicate there may be a connection between these two insulating states\cite{velasco_jr_transport_2012, freitag_spontaneously_2012}.

The $\nu=0$ state in BLG occurs at half filling of the lowest Landau level. Although the BLG zero-energy Landau level has an additional orbital degeneracy stemming from two accidentally degenerate magnetic oscillator wavefunctions\cite{mccann_landau-level_2006}, minimization of exchange energy favors singlet pairs in the orbital space at all even filling factors~\cite{barlas_intra-landau-level_2008}. This leaves an approximate SU(4) spin and pseudospin symmetry for the remaining ordering of the ground state. Comparable ground state competition has also been studied in double quantum wells in GaAs 2-dimensional electron systems (2DESs)~\cite{das_sarma_double-layer_1997}. Quantum phase transitions, in particular from canted antiferromagnetic to ferromagnetic ordering, can occur in those systems as an in-plane magnetic field is applied (Zeeman energy is increased)~\cite{pellegrini_evidence_1998, grivei_multiple_2003}.  

    Our experiment is carried out in BLG samples with top and bottom gates in which thin single crystal hBN serves as a high quality dielectric on both sides (Fig. 1 bottom inset). By controlling top gate voltage ($V_{TG}$) and bottom gate voltage ($V_{BG}$) we can adjust carrier density $n$ and perpendicular electric displacement field $D$ independently. Additionally, tilting the sample in the magnetic field allows us to independently control the Coulomb energy in a Landau level, $e^2/\ell_B$, where the magnetic length $\ell_B$ is determined by the perpendicular magnetic field $B_\perp$, and Zeeman energy (determined by the total magnetic field $B_{tot}$). Since in BLG $B_{tot}$ and $D$ separately couple to spin and pseudospin, respectively, our experimental setup allows us to tune anisotropies and characterize the broken-symmetry QH states in the approximate SU(4) spin-pseudospin space (see supplementary information).

    Fig. 1 shows the resistance measured as a function of the two gate voltages. At zero magnetic field, the resistance at the CNP depends monotonically on $|D|$ due to the opening of a single particle energy gap \cite{mccann_asymmetry_2006, taychatanapat_electronic_2010}. With increasing applied magnetic field (Fig. 1 top inset) our device shows the same behavior reported elsewhere\cite{weitz_broken-symmetry_2010, velasco_jr_transport_2012, kim_spin-polarized_2011}, with a layer-polarized (LP) insulator (which we denote phase (i)) observed at large applied displacement field, separated from a low displacement field and high magnetic field quantum Hall insulator (phase (ii)) by a well defined boundary of finite conductance. For the remainder of the paper we focus on the behavior of phase (ii) as we vary the relative Zeeman versus Coulomb energy scales in a rotated field geometry.  
   
\begin{figure}[t]
    \begin{center}
    \includegraphics[width=\linewidth]{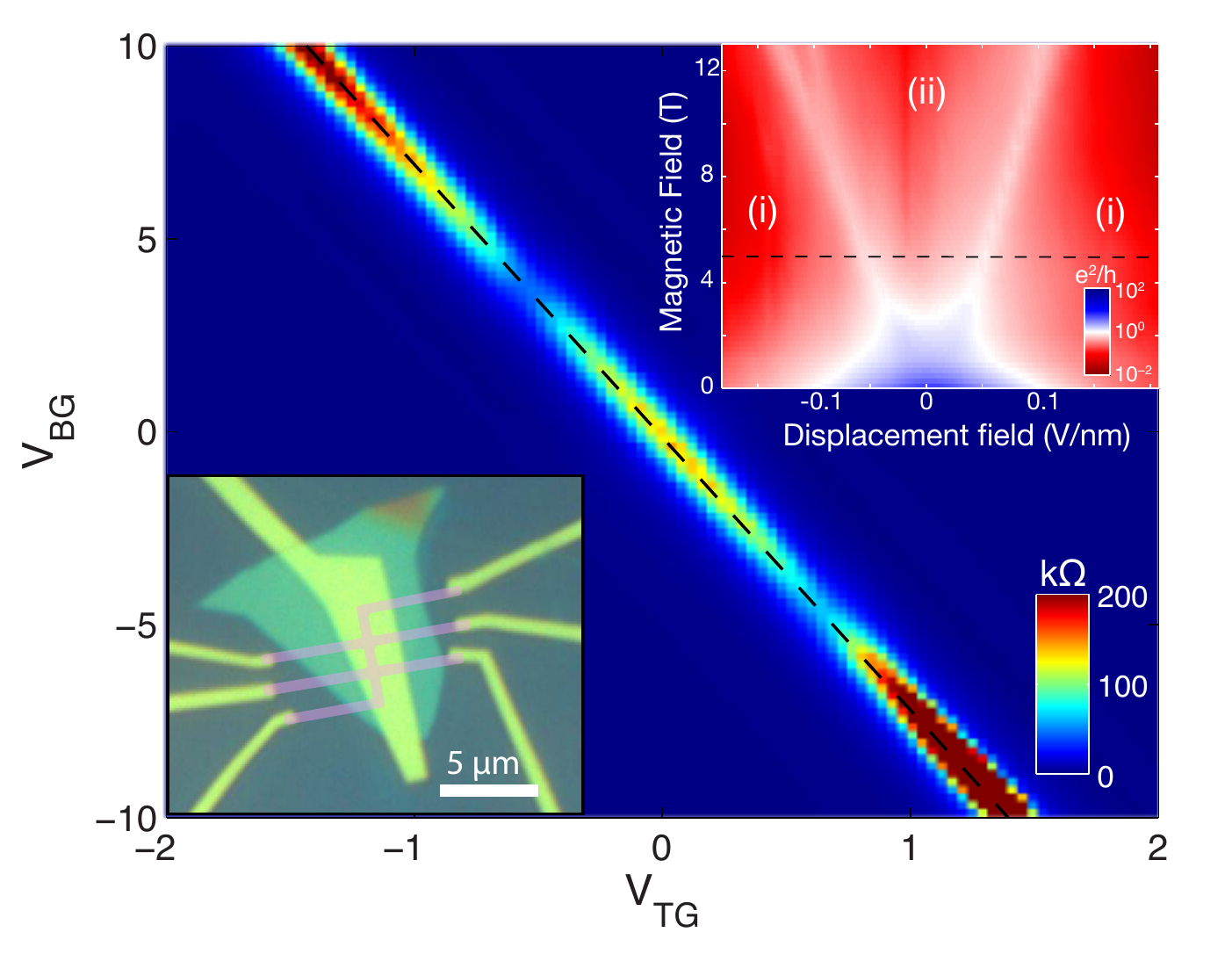}
        \caption{\textbf{Device summary.} Plot of $R_{xx}$ versus top gate voltage $V_{TG}$ and bottom gate voltage $V_{BG}$ at $B_\perp=B_{tot}=5$ T. Data taken at 1.5 K. Dashed line indicates the CNP. Top right inset shows four-terminal conductance at the CNP plotted against $B_\perp$ and $D$. Dashed line in inset corresponds to dashed line in main plot. Bottom left inset shows an optical image of the device used in this experiment with false coloring of the graphene (pink).}
		\label{fig1}
	\end{center}
\end{figure}

    Fig. 2a shows a series of conductance maps in the vicinity of the CNP. Each pane corresponds to a fixed perpendicular applied magnetic field of 1.75 T, but with varying total field. At a total field of 1.75~T (that is, $B_\perp=B_{tot}$), the distinct insulating regimes at low and high $|D|$ are clearly visible, as discussed in Fig. 1. At low $|D|$ the four-terminal conductance increases dramatically with increasing total magnetic field $B_{tot}$, reaching a value of around $3-4~e^2/h$, after which it becomes relatively insensitive to either displacement field or magnetic field (Fig. 2b). This rapidly increasing behavior of conductance at fixed $B_\perp$ is consistent with the gradual closing of a gap in phase (ii) as $B_{tot}$ increases. At fixed Coulomb energy, this decreasing gap with increasing Zeeman energy suggests that phase (ii) corresponds to a spin-unpolarized ground state\cite{feldman_broken-symmetry_2009, zhao_symmetry_2010, veligura_transport_2012, kharitonov_canted_2012}. In addition, the saturation of conductance at a metallic value around $\sim 4~e^2/h$ at $B_{tot}\gtrsim 15$~T strongly suggests a transition into a new phase with a fully closed transport gap, which we label phase (iii).

    The transition between insulating and metallic behaviors can further be confirmed by investigating the temperature dependence of the conductance. Fig. 2 c and d plot four-terminal conductance $G$ at the CNP against $D$ at fixed $B_{\perp}=2.5$~T with $B_{tot}=2.5$~T and 13.1~T, respectively. Temperature $T$ varies between 3-18~K. At small $|D|$, $G$ exhibits weakly insulating behavior for $B_{tot}=B_\perp$, while at $B_{tot}\gg B_\perp$, metallic temperature dependence (increasing $G$ with decreasing $T$) is observed. Fig. 2e further contrasts the metallic versus insulating classification in different phases by plotting conductance at the CNP against temperature for both large $D$ and $D=0$. $G(T)$ shows insulating behavior for phase (ii) ($B_\perp \approx B_{tot}$) and metallic behavior for phase (iii) ($B_\perp \ll B_{tot}$). At intermediate values of $B_{tot}$ the conductance is relatively temperature insensitive, suggesting a gradual crossover between the two phases. Interestingly, the cusps of high conductance that mark the transition between the LP and non-LP phase show metallic temperature dependence even in a fully perpendicular field. At high $D$ (i.e. in phase (i)), the conductance remains relatively insensitive to the applied in-plane magnetic field. This is expected for a LP $\nu=0$ QH state, where the energy gap is determined by the applied displacement field.

\begin{figure}[t]
    \begin{center}
    \includegraphics[width=\linewidth]{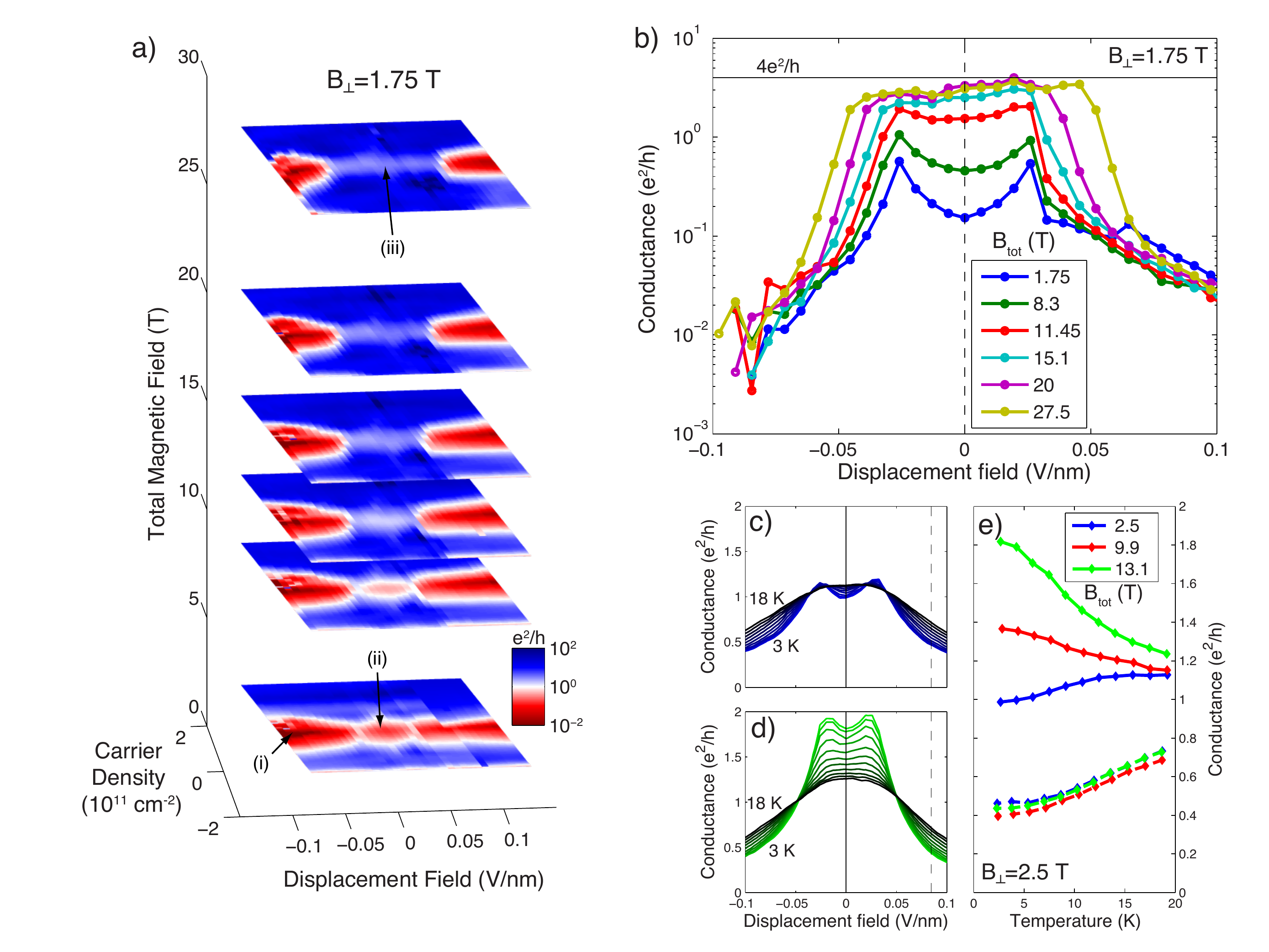}
		\caption{\textbf{Tilted field transport.} \textbf{a)} $1/R_{xx}$ plotted against carrier density, displacement field, and total magnetic field with $B_\perp$ fixed at 1.75 T. Data taken at 350 mK. \textbf{b)} Four-terminal conductance at the CNP ($G=1/R_{xx}$) plotted against displacement field $D$ for a variety of total magnetic fields, $B_\perp$ fixed at 1.75 T. Dashed line indicates the zero displacement field points plotted in c). Solid line marks the theoretically expected $4e^2/h$ conductance for the F phase. Data taken at 350 mK. \textbf{c-d)} Temperature dependent four-terminal conductance at the CNP plotted against displacement field, with $B_\perp=2.5~\mathrm{T}$, $B_{tot}=$2.5 T (c), 13.1 T (d). Colored curves are taken at 3 K, black curves at 18 K, and intermediate traces are spaced by 1.5 K. Dashed (solid) vertical lines at a fixed displacement field correspond to dashed (solid) lines in e). \textbf{e)} Four-terminal conductance at the CNP for different total magnetic fields plotted against temperature, all at $B_\perp=2.5~\mathrm{T}$. Solid lines demonstrate how behavior changes as the total field is increased for zero displacement field. Dashed lines demonstrate similar behavior for all total magnetic fields at $D=0.085$ V/nm.}
		\label{fig2}
	\end{center}
\end{figure}

\begin{figure}[t]
    \begin{center}
	\includegraphics[width=\linewidth]{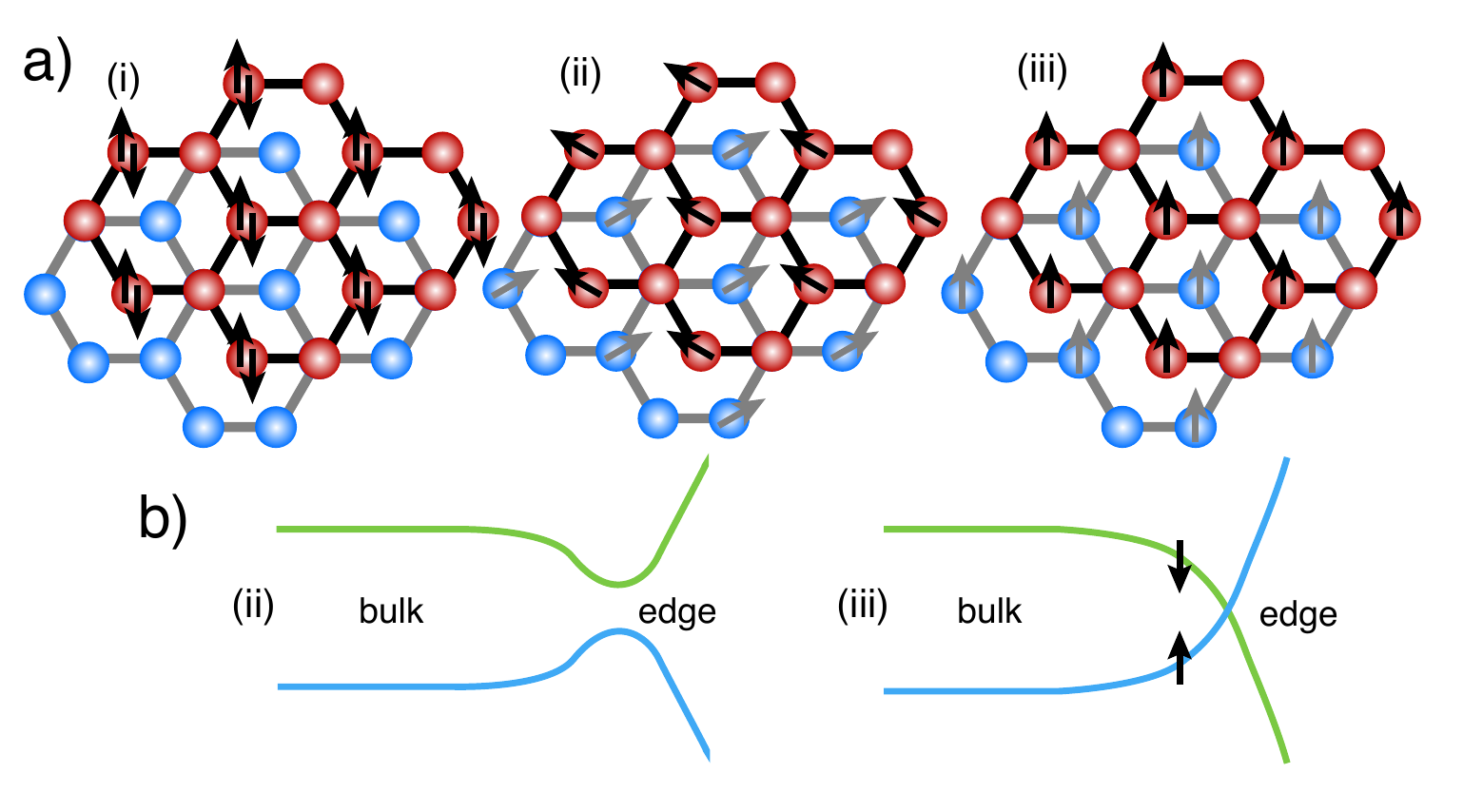}
		\caption{\textbf{Candidate states.} \textbf{a)} Spin and pseudospin configurations in the BLG $\nu=0$ state. Cartoons represent the layer polarized (LP) (i), canted antiferromagnet (CAF) (ii), and ferromagnet (F) (iii) phases. \textbf{b)} Diagram of predicted bulk and edge energy levels for the CAF (ii) and F (iii) phases.}
		\label{fig3}
	\end{center}
\end{figure}

A general theoretical analysis~\cite{kharitonov_canted_2012} proposes four possible ground states for $\nu=0$ in BLG: partially layer-polarized (PLP), fully layer-polarized (FLP), canted antiferromagnet (CAF), or ferromagnet (F). Considering that layer pseudospin is synonymous with sublattice and valley in the lowest Landau level~\cite{mccann_landau-level_2006}, these four different scenarios can be visualized in terms of spin and sublattice of BLG (Fig. 3a). Which state the system favors depends on both anisotropies in the electron-electron and electron-phonon interactions, and external symmetry-breaking terms like Zeeman splitting $\epsilon_Z$ and interlayer potential energy $\epsilon_V$. The PLP/FLP phases are favored in the large $\epsilon_V$ limit, and the F phase is favored in the large $\epsilon_Z$ limit. Unlike in GaAs double wells, these ground states may be accompanied by dramatically different transport signatures. The BLG ground state configurations discussed above are all predicted to be insulators except for the F phase. For the F phase, the highest filled band acquires an electron-like dispersion at the edge while the lowest unoccupied band acquires a hole-like dispersion (Fig. 3b)\cite{kharitonov_edge_2012}. This forces a level crossing at the edge which should result in metallic conductance of $4e^2/h$ due to spin-polarized counter-propagating edge states\cite{abanin_spin-filtered_2006, fertig_luttinger_2006, kharitonov_edge_2012}. Interpreting our results in this framework, we can identify phase (i) with either the PLP or FLP state, phase (ii) with the CAF state, and phase (iii) with the F state. 

We now focus on the phase diagram of the different ground states classified above. The phase transitions between the LP, CAF, and F phases can be mapped out by locating the boundary between the metallic ($G \gtrsim e^2/h$) and insulating ($G \ll e^2/h$) regimes as a function of $D$ and $B_{tot}$ with fixed $B_\perp$. For a given $B_{tot}$, we can define the critical displacement field for transition $D^c$ as the $D$ value that gives maximum change of $\partial G/\partial D$ in $G(D,B_{tot})$ at fixed $B_\perp$. Fig. 4a shows the values of $D^c$ and $B_{tot}$  extracted from Fig. 2b ($B_\perp=1.75$~T). In this $D$-$B_{tot}$ map three different phases can be identified, corresponding to the LP, CAF, and F phases. Note that the transition between the CAF and LP phases is sharp, as the conductance changes by an order of magnitude over a $D$ variation $\Delta D \lesssim 0.02$ V/nm. By contrast the transition from CAF to F is gradual. This observation is consistent with the theoretical expectation that the LP to CAF or F transition is first order while the CAF to F transition is second order\cite{kharitonov_canted_2012}.

\begin{figure*}[t]
    \begin{center}
    \includegraphics[width=\linewidth]{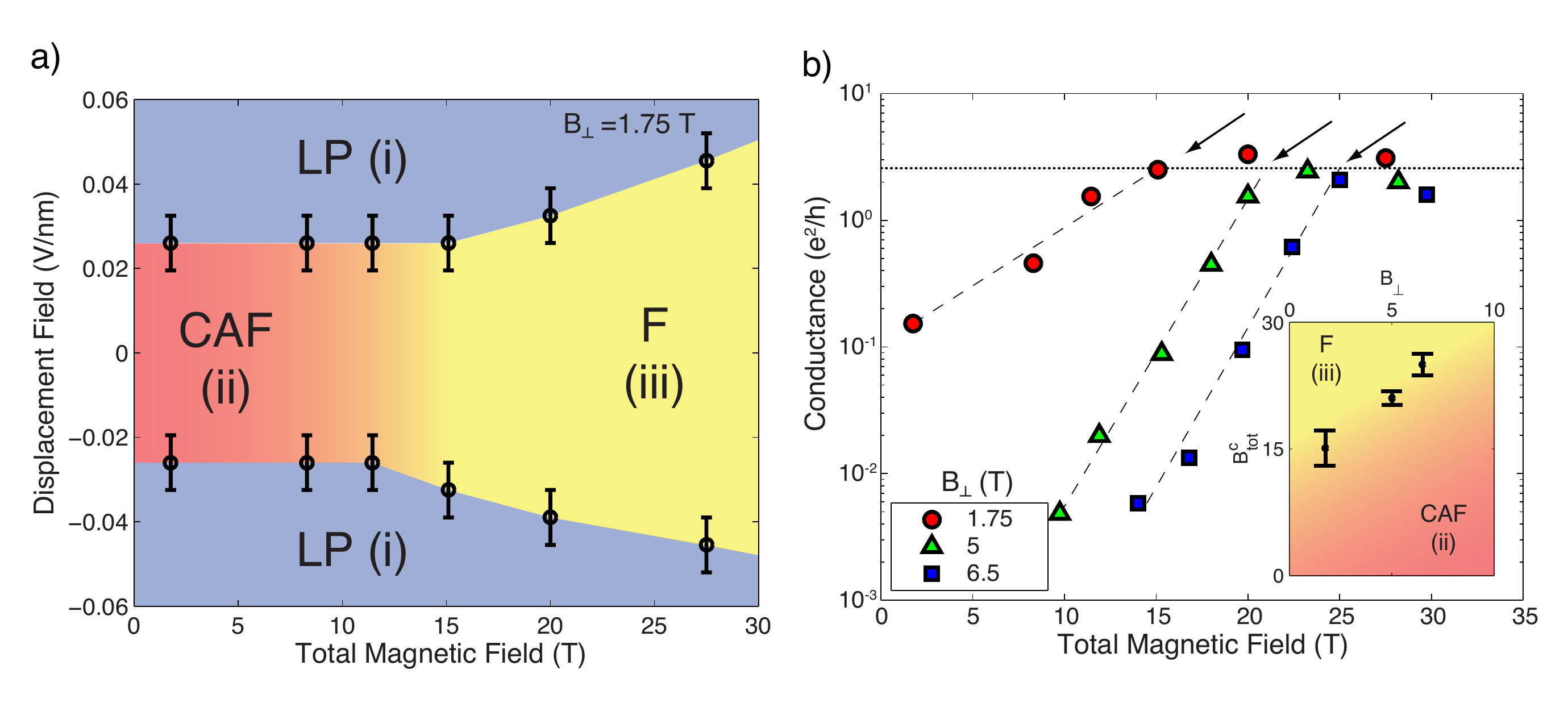}
		\caption{\textbf{Phase Transitions.} \textbf{a)} Different phases of the $\nu=0$ state mapped as a function of $D$ and $B_{tot}$ at $B_\perp=1.75~\mathrm{T}$. At low $B_{tot}$, the phase boundary between LP and CAF is determined by the conductance spike separating the two insulating phases. At large $B_{tot}$, the phase boundary between LP and F is determined by the point where conductance begins to exponentially decrease with displacement field. The gradual transition between CAF and F with increasing $B_{tot}$ is marked by the gradual color change, with F stabilizing at high $B_{tot}$. \textbf{b)} Four-terminal conductance at the CNP with $D=0$, plotted against $B_{tot}$ for three different values of $B_\perp$. Dashed lines indicate regions of exponential increase of conductance and the dotted line marks saturated conductance. The critical magnetic field $B_{tot}^c$ for the phase transition is marked by arrows. Inset shows $B_{tot}^c$ as a function of $B_\perp$ separating the CAF and F phases in a $B_{tot}$ versus $B_\perp$ parameter space.}
		\label{fig4}
	\end{center}
\end{figure*}

By identifying characteristic energy scales for the external symmetry breaking terms, we can comment quantitatively on the phase boundaries for a fixed $B_\perp$. We identify Zeeman energy $\epsilon_Z=g\mu_B B_{tot}$, where $g$ and $\mu_B$ are the electron spin g-factor and the Bohr magneton respectively, and the interlayer potential energy $\epsilon_V= Dde/2$, where $d$ is the interlayer distance in BLG. For $B_\perp=1.75$ T, at low $B_{tot}$ the CAF-LP transition occurs at $\epsilon_V\approx5$ meV, and at $\epsilon_V=0$ the CAF-F transition occurs at $\epsilon_Z\approx 0.8$ meV. The F-LP transition occurs along a line whose slope $\epsilon_V/\epsilon_Z\approx2$. The fact that $\epsilon_V/\epsilon_Z$ at the F-LP phase boundary is of order unity is consistent with the increasing stability of a spin-polarized state with increasing magnetic field, providing further confirmation that the conducting $\nu=0$ phase is a ferromagnet. This observation is in qualitative agreement with theoretical predictions\cite{kharitonov_canted_2012}.

Since at the CNP and $D=0$, $B_\perp$ is the only physical quantity to control the strength of e-e interactions, the phase boundary between the CAF and F phases can be further investigated by observing $G$ as a function of $B_{tot}$ with different values of fixed $B_\perp$ (see supplementary information). Fig. 4b shows $G(B_{tot})$ for three different fixed values of $B_\perp$ at $D=0$. Generally, the behavior of $G$ can be characterized by two different regimes. For $B_{tot} \sim B_\perp$ (i.e. the CAF phase), $G$ increases exponentially as $B_{tot}$ increases. For $B_{tot} \gg B_\perp$ (i.e. the F phase), $G$ is insensitive to $B_{tot}$, saturating close to $4e^2/h$. We define the critical magnetic field $B_{tot}^c$ which separates these two regimes by the intersection of guide lines for the exponential and saturated regimes, as marked by the arrows in Fig. 3b. The identification of $B_{tot}^c$ at different $B_\perp$ allows us to map the phase boundary between CAF and F as a function of $B_\perp$. The inset of Fig. 4b displays $B_{tot}^c$ as a function of $B_\perp$. Although there are only a few data points available to indicate the phase boundary, the general trend of the CAF-F phase transition can be inferred from them. 

Finally, we remark on the similarity between the F phase of BLG at $\nu=0$ and the quantum spin Hall effect (QSHE) in HgTe\cite{bernevig_quantum_2006, konig_quantum_2007}. In the QSH phase quantized conduction is due to spin-polarized counter-propagating edge states, while the bulk remains incompressible. An analogous scenario is expected for BLG in the F quantum Hall state\cite{abanin_spin-filtered_2006, fertig_luttinger_2006, kharitonov_edge_2012}, except that BLG carriers an additional orbital degeneracy.  The F state in BL graphene is therefore expected to exhibit a four terminal conductance of $4e^2/h$, compared with $2e^2/h$ for HgTe. Similar to reported measurements of HgTe\cite{konig_quantum_2007, roth_nonlocal_2009}, our devices show imperfect quantization with measured conductance values less than $4e^2/h$ in the F phase.  The reduced conductance may be due to backscattering in the edge state or through remaining impurity states in the bulk. Unambiguous experimental support for the existence of spin-polarized counter-propagating edge states should be provided by well-quantized nonlocal measurements \cite{roth_nonlocal_2009} (see supplementary information). 

    In conclusion, we have measured dual-gated graphene bilayers at the $\nu=0$ state. In a perpendicular magnetic field, we observe a transition between two incompressible states at a finite displacement field, indicating that the $D=0$ state is not layer polarized. At low displacement fields, we observe four-terminal conductance increase and then saturate as in-plane magnetic field increases, showing that the $D=0$, $B_{tot}=B_\perp$ state is also not spin polarized. This is consistent with a quantum phase transition driven by in-plane magnetic field between the canted antiferromagnet quantum Hall state and the ferromagnet. The near quantization of the 4-terminal conductance to $4e^2/h$ agrees with predictions that the ferromagnetic quantum Hall ground state carries spin-polarized counter-propagating edge states.

\section{Methods}
The device we report data from was fabricated with mechanically exfoliated graphene and hBN using a polymer membrane transfer process described in previous work~\cite{dean_boron_2010}. The graphene was etched into a Hall bar pattern prior to deposition of the top hBN layer. The bottom gate is a global silicon gate, and both the contacts and the top gate are Cr/Pd/Au stacks of thickness 1/10/50 nm. Three devices fabricated in a similar way exhibited similar data trends. We present data sets obtained from the highest quality device.

These devices were measured either in a sample-in-$^4$He-vapor variable temperature cryostat with 14~T magnet in house or in a sample-in-$^3$He-vapor cryostat, mounted in the bore of a 31~T resistive magnet at the National High Magnetic Field Lab in Tallahassee, FL. Measurements were performed using a 1 mV voltage bias. The numerous features present in a gate voltage trace allowed precise angle calibration, with $B_\perp$ determined to better than 1\% accuracy. The longitudinal resistance $R_{xx}$ and Hall resistance $R_{xy}$ are measured in a four-terminal geometry, so that contact resistances can be excluded. Four-terminal conductance at the CNP is obtained by taking $G=1/R_{xx}$.
 
\bigskip

\section{Acknowledgements}

The authors thank Maxim Kharitonov for useful discussions.
Portions of this experiment were conducted at the National High Magnetic Field Laboratory, which is supported by National Science Foundation Cooperative Agreement No. DMR-0654118, the State of Florida and the US Department of Energy. We thank S. Hannahs, T. Murphy, and A. Suslov for experimental assistance at NHMFL. This work is supported by AFOSR MURI. PM acknowledges support from ONR MURI and FENA. AFY and PK acknowledge support from DOE (DE-FG02-05ER46215) for carrying out experiments and INDEX for sample fabrication.

\section{Contributions}

PM, CRD, and AFY designed and conceived the experiment. TT and KW synthesized hBN samples, PM fabricated the samples. PM, CRD, and AFY performed the measurements. PM, CRD, and PK analyzed the data and wrote the paper. JH, KS, and PK advised on experiments.


\end{document}